\documentclass[11pt]{article}

\usepackage{cite}
\usepackage{amsmath}
\usepackage{amsfonts}
\usepackage{epsf}

\def\lfrac#1#2{{}^{#1\!}\kern-.0em/_{#2}}
\def\buildrel#1\under#2{\mathrel{\mathop{\kern0pt #2}\limits_{#1}}}

\def\bbox#1{\mbox{\boldmath$#1$}}

\begin{document}

\null \hfill physics/0110053\\
\vspace*{1.5cm}

\begin{center}
{\Large \sf Nonresonant Effects in One-- and Two--Photon Transitions}
\end{center}

\vspace{0.05cm}

\begin{center}
Ulrich~D.~Jentschura$^{a),b)}$ and Peter J. Mohr$^{b)}$
\end{center}

\vspace{0.05cm}

\begin{center}
$^{a)}${\it Institute of Theoretical Physics,}\\
{\it Dresden University of Technology, 01062 Dresden,
Germany}\\[1ex]
$^{b)}${\it National Institute of Standards and Technology,\\
Mail Stop 8401, Gaithersburg, Maryland 20899-8401, USA}
\end{center}

\vspace{0.6cm}

\begin{center}
\begin{minipage}{10.5cm}
{\underline{Abstract}}
We investigate nonresonant contributions to resonant Rayleigh 
scattering cross sections of atoms. The problematic nonresonant
contributions set a limit to the accuracy to which
atomic spectra determine energy levels. We discuss the 
off-resonance effects in one-photon transitions. We 
also show that off-resonance contributions for the 1S--2S two-photon
transition in atomic hydrogen are negligible at current and projected
levels of experimental accuracy. The possibility of a differential
measurement for the detection of off-resonance effects in one-photon
transitions in atomic hydrogen is discussed.
\end{minipage}
\end{center}

\vspace{0.6cm}

quantum electrodynamics -- specific calculations\\
\vfill
\begin{center}
\begin{minipage}{14cm}
\begin{center}
\hrule
{\bf \scriptsize
\noindent electronic mail: 
ulj@nist.gov, mohr@nist.gov}
\end{center}
\end{minipage}
\end{center}

\newpage

%
%
\section{Introduction}

Recently, the dramatic progress in laser-spectroscopic experiments
in atomic hydrogen~\cite{NiEtAl2000} has sparked interest in theoretical 
calculations at highly improved 
accuracy~\cite{JeMoSo1999,MeRi2000,Ye2000,JeMoSo2001pra}. 
This raises interesting questions regarding the relation 
of the resonance peak in the scattering cross section, which is observed in 
experiments, and the actual difference in the real parts 
of the energies of the two levels involved in the atomic 
transition\footnote{in constrast to the ``real part'', the ``imaginary part'' 
of the energy $-{\mathrm i}\,\Gamma/2$
corresponds to the ``radiative width'' of
the state which is related to the ``radiative lifetime'' $\tau$ according to
$1/\tau = \Gamma/\hbar$}.
In short, one may ask to which level of
accuracy atomic spectra determine energy levels.
Related questions are of prime importance 
for the determination of fundamental
constants~\cite{MoTa2000}.

\begin{figure}[htb]
\begin{center}
\begin{minipage}{12cm}
\centerline{\mbox{\epsfysize=5.0cm\epsffile{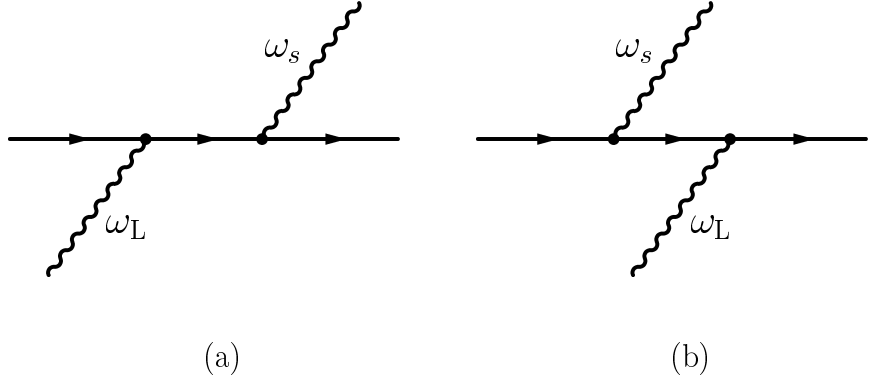}}\hbox to 0.75in{}}
\caption{\label{fig1} A two-photon process with absorption
and emission.  Time increases from left to right.
The atom absorbs 
one laser photon with frequency $\omega_{\mathrm L}$ and
emits one photon with frequency $\omega_{\mathrm s}$. 
The electron propagator is that of the 
bound electron, which is assumed to be in its ground state
in both the initial and final configuration.}
\end{minipage}
\end{center}
\end{figure}

\begin{figure}[htb]
\begin{center}
\begin{minipage}{12cm}
\centerline{\mbox{\epsfysize=4.0cm\epsffile{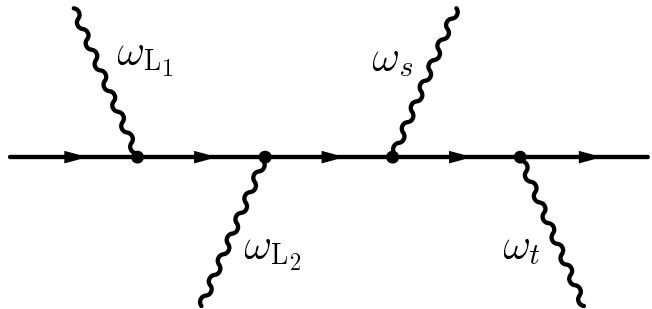}}\hbox to 0.75in {}}
\caption{\label{fig2} A four-photon process. Two laser photons
with frequency $\omega_{\mathrm L_1}$ and $\omega_{\mathrm L_2}$ are absorbed,
and two photons with frequencies $\omega_{s,t}$ are emitted.  
In many experiments, $\omega_{\mathrm L} = \omega_{\mathrm L_1} = 
\omega_{\mathrm L_2}$. As in Fig.~\protect{\ref{fig1}}, the cross section 
is given by the sum over permutations
of the laser photons and the emitted photons.
The resonance condition $2\omega_{\mathrm L} = E_2 - E_1$ corresponds to
the absorption of two photons.}
\end{minipage}
\end{center}
\end{figure}

These issues are most easily dealt with if one assumes that the 
scattering process is described to a good approximation by a 
Kramers--Heisenberg formula~\cite{KrHe1925}. 
For {\em one-photon} transitions,
the excitation of the atom from the ground state
by a laser photon and photon
emission\footnote{This process is an elastic 
scattering process (the atom returns to the ground state) 
that contributes to the Rayleigh scattering cross section. 
This is in contrast to inelastic scattering of highly
energetic photons by free or ``approximately free''
electrons (Compton scattering) and inelastic scattering in
the low-frequency domain (Raman scattering); see, e.g., the 
discussion on pp.~374--376 of~\cite{Lo2000}.} 
is well described by the two diagrams in
Fig.~\ref{fig1}. For {\em two-photon} transitions, the situation is more
involved. We consider here a process where
two interactions with laser-photons are accompanied by
the emission of two photons (see Fig.~\ref{fig2}). 

It has been pointed out as early as 1952~\cite{Lo1952}, that the
experimental spectrum of atomic hydrogen does not reproduce
the energy level differences precisely, and that nonresonant
contributions to photon scattering shift the observed resonance 
peaks relative to the energy level differences
by a frequency $\delta \omega$ which is of the order of
\begin{equation}
\label{LowEstim}
\delta \omega \sim \alpha^2 \, (Z\alpha)^6 \, \frac{m c^2}{\hbar}\,,
\end{equation}
where $m$ is the electron mass, $\alpha$ is the fine structure constant,
$Z$ the nuclear charge number, and $c$ the speed of light; $\hbar = h/(2\pi)$ 
is the natural unit of action where $h$ denotes Planck's constant. 
This shift $\delta \omega$ occurs for 
one-photon transitions where the atom returns to its ground state
after the photon emission. Later, the interesting fact that resonance   
peaks do not necessarily determine atomic energy levels was 
discussed in Ref.~\cite{Bi1995}. We also refer to the related 
investigations~\cite{LundeenThesis,La1983,LaKlDm1993,KaEtAl1992,LaEtAl1993,
LaKaGo1994,LaSoPlSo2001}.
In a recent theoretical calculation of higher-order
radiative effects for hydrogenic P states~\cite{JeSoMo1997},
it has also been pointed out that an accurate analysis of the 
line shape is necessary at the level of the current 
theoretical uncertainty. 

{\em A priori}, the detection of 
a nonresonant contribution of the order of~(\ref{LowEstim}) in a
one-photon transition would require the determination of the 
peak of a hydrogenic transition frequency with an
uncertainty of roughly $10^{-7}$ relative to the width of the transition
frequency [the natural radiative 
decay rate $\Gamma$
of a typical bound state in atomic hydrogen is of the order
of $\alpha\,(Z\alpha)^4$ in units of $m\,c^2/\hbar$,
and we consider the quantity $\delta \omega/\Gamma$].  
In Sec.~\ref{Differential}, we will consider a
differential cross section,
in which the nonresonant contribution could be observed
more clearly.

%
%

%
%

%
%
\begin{figure}[htb]
\begin{center}
\begin{minipage}{12cm}
\centerline{\mbox{\epsfysize=4.0cm\epsffile{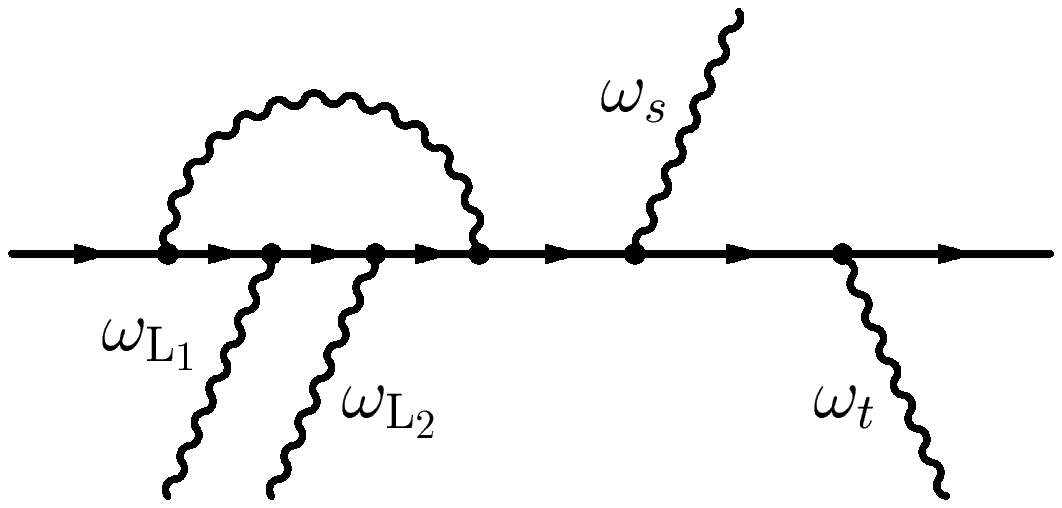}}\hbox to 0.75in {}}
\caption{\label{fig3} A four-photon process with a vertex
correction. 
The vertex corrections are expected to give corrections of relative order
$\alpha(Z\alpha)^2$ to the cross section.}
\end{minipage}
\end{center}
\end{figure}

This paper is organized as follows:  In Sec.~\ref{TwoPhoton},
we estimate nonresonant frequency shifts in both one- and 
two-photon transitions with special attention devoted to the 1S--2S two-photon
transition in hydrogen (our estimates are also of relevance 
for other two-photon processes). We do not consider the effect of
radiative vertex corrections to the photon absorption and emission
on the line shape in this paper.  In Sec.~\ref{Differential},
we consider a differential cross section measurement in which nonresonant 
effects could be observed with optical transition
frequencies in atomic hydrogen. Conclusions are left to
Sec~\ref{Conclusions}.

%
%
\section{One-- and Two--Photon Transitions: Off--Resonant Effects}
\label{TwoPhoton}

The first theoretical description of atomic two-photon transitions 
was given in Ref.~\cite{GM1931}, with the first experimental observation
following in Ref.~\cite{KaGa1961}. The description becomes less involved
if one assumes that both the initial and the final state of the 
transition are ``asymptotic states,'' i.e.,~states which can be 
used as initial and final states for $S$-matrix elements. This 
is a valid approximation only in the case of a long radiative 
lifetime for both the initial and the final state of the transition
(of course, only the ground state will survive in the asymptotic limit of 
infinitely large time.)

We consider, within the ``asymptotic-state'' approximation,
the differential cross section for a two-photon excitation 
process $|1\rangle \to |f\rangle$ through an intermediate
state $|i\rangle$ (here, we assume the state $|1\rangle$ to be the ground 
state). We restrict the discussion to a situation
in which two photons, one from each of two counter-propagating 
laser beams are absorbed by the atom. 
This eliminates first-order Doppler shifts
$\delta \omega_{\mathrm D} = \pm \bbox{k} \cdot \bbox{v}$ 
and corresponds to the situation encountered in current 
high-precision experiments on S $\to$ S transitions
in atomic hydrogen~\cite{NiEtAl2000}, where 
(first-order) Doppler-broadened background is suppressed.
The cross section in this case is proportional to
\begin{equation}
\label{AsympApprox}
\left| \sum_i
\frac{\left(\bbox{\varepsilon}_\lambda \cdot \bbox{D}_{fi} \right)
\, \left(\bbox{\varepsilon}_\lambda \cdot \bbox{D}_{i1}\right)} 
  {E_i - E_1 - \hbar \omega_{\mathrm L}} \right|^2 \,
\delta (E_f - E_i -2\hbar \omega_{\mathrm L}) \,,
\end{equation}
where the sum over $i$ runs over all intermediate states,
with the appropriate mutliplicity for degenerate states.
The polarization vector 
of the laser beam is $\bbox{\varepsilon}_\lambda$;
$\omega_{\mathrm L}$ is the laser frequency and
$\bbox{D}_{ij} = \langle i | e\,\bbox{x} | j \rangle$
is the dipole matrix element
\footnote{If Doppler-broadened background is not fully suppressed,
then the Doppler-free cross section (\ref{AsympApprox}) manifests
itself as a narrow peak which is superimposed on 
Doppler-broadened background. In this case,
the Doppler-broadened background must be taken into account
[see Eq.~(7.43) of~\cite{De1993}] via an additional
additive contribution in (\ref{AsympApprox}) involving
a factor $\exp\left[-(E_f - E_1 - 2 \hbar \omega_{\mathrm L})^2/
(2 k_{\mathrm L} v_{\mathrm w})^2 \right]$,
where $v_{\mathrm w} = \sqrt{2 k_{\mathrm B} T/m}$ is the thermal
velocity of the atoms (at the peak of the
thermal probability distribution), and 
$k_{\mathrm L} = \omega_{\mathrm L}/c$.}.

As is evident from Eq.~(\ref{AsympApprox}),
for two-photon transitions, we can expect large resonant
effects if the light of one of the lasers excites an 
intermediate resonance ($\hbar \omega_{\mathrm L} \approx 
E_i - E_1$). In this case, the energy denominator in
(\ref{AsympApprox}) becomes small (we would then have to
include imaginary parts according to $E_i \to
E_i - {\mathrm i} \Gamma_i/2$)
and the cross section peaks. Such an ``intermediate-state'' resonance
occurs only if there exists an intermediate state with the 
appropriate energy.

For the current high-precision hydrogen measurements~\cite{NiEtAl2000},
all energy denominators in the summation in (\ref{AsympApprox})
are non-vanishing even at the peak of the two-photon resonance,
which is near $\omega_{\mathrm L} \approx (E_f - E_1)/(2\hbar)$.
A possible exception would be the case of an atomic system
in which there is accidentally an intermediate state 
present with an energy midway between $E_1$ and $E_f$,
that is $E_i \approx (E_f - E_1)/2$, but in
general, this is {\em not} relevant to atomic hydrogen,
because the spectrum is not equally spaced.

Although the Doppler-free two-photon 1S--2S resonance
is extremely narrow, with a natural line width 
of $\Gamma_f = \Gamma_{\mathrm{2S}}$
(we recall that the natural width of the 2S level 
in atomic hydrogen is 1.3~Hz~\cite{MaMo1978}),
nonresonant contributions need to be estimated in an improved 
framework that avoids the ``asymptotic-state'' approximation
inherent in Eq.~(\ref{AsympApprox}).  To do this,
we first consider a two-photon process with one photon being
absorbed and one being emitted (see Fig.~\ref{fig1}), as 
discussed by Low~\cite{Lo1952}.  This approach has recently
been applied in a study of the line shape in 
Ref.~\cite{LaSoPlSo2001}.
The atom is excited from its ground state $|1\rangle$ to an
intermediate state $|i\rangle$ by interaction with a laser
of frequency $\omega_{\mathrm L}$ and returns to the ground 
state $|1^\prime\rangle$ via spontaneous emission. 
The differential cross section
is proportional to~\cite{KrHe1925,Lo1952}
\begin{equation}
\label{TwoPhotonCross}
\frac{{\mathrm d}\sigma}{{\mathrm d}{\it \Omega}}
(\bbox{\varepsilon}_{\mathrm L}, 
\bbox{\varepsilon}_{\mathrm s}) 
\propto \left| \sum_{i}
\frac{\left(\bbox{\varepsilon}^*_{\mathrm s} 
  \cdot \bbox{D}_{1^\prime i}\right) \,
  \left(\bbox{\varepsilon}_{\mathrm L} \cdot
  \bbox{D}_{i1}\right)}{E_i - {\mathrm i}\, 
  \Gamma_i/2 - (E_1 + \hbar \omega_{\mathrm L})} 
+ \sum_{i}
  \frac{\left(\bbox{\varepsilon}_{\mathrm L} \cdot 
  \bbox{D}_{1^\prime i}\right) \,
  \left(\bbox{\varepsilon}^*_{\mathrm s} \cdot
  \bbox{D}_{i1}\right)}{E_i - (E_1 - \hbar \omega_{\mathrm L})}
   \right|^2\,,
\end{equation}
where $\bbox{D}_{ij} = \langle i | e\, \bbox{x} | j \rangle$ is
again the dipole matrix element, and $\bbox{\varepsilon}_{\mathrm s}$
and $\bbox{\varepsilon}_{\mathrm L}$ are the polarization vectors
of the emitted photon and the laser photon, respectively.
We work within the relativistic dipole approximation in which the 
replacement 
$\langle i | \bbox{\alpha}\, {\rm exp}({{\rm i}\,\bbox{k} \cdot \bbox{x}}) 
| j \rangle \rightarrow
\langle i | \bbox{\alpha} | j \rangle =
{\rm i} \, (E_i - E_j)\langle i | \bbox{x} | j \rangle$
is made for the transition matrix element. As it will become clear
in the sequel, it is necessary to retain $E_i$ and $E_j$ as 
relativistic energies, including fine-structure effects.
The total cross section is obtained by summing over final angular
momentum states
of the electron, summing over the polarization states
of the emitted photon, and integrating over directions of the emitted
photon.  If the hydrogen atoms in the initial state are unpolarized,
the cross section is also averaged over initial angular momentum
states.  

The two terms in the sum
on the right-hand side correspond to Figs.~\ref{fig1}(a) and
(b), respectively. If desired,
missing prefactors in (\ref{TwoPhotonCross}) can be
restored according to Eq.~(8.7.6) of Ref.~\cite{Lo2000}.
This factor includes a dependence on the laser frequency
$\omega_{\mathrm L}$ and on the frequency of the emitted
photon $\omega_{\mathrm s}$ of the form $\omega_{\mathrm L} \,
\omega_{\mathrm s}^3$. In this article, we do not consider frequency
shifts related to this additional frequency dependence of the cross
section; these effects are already present in the resonance
approximation and are included in the basic line shape through a
multiplicative factor.

In contrast to the two-photon absorption
described by Eq.~(\ref{AsympApprox}), we
now have {\em two} terms in the amplitude for the
emission-absorption process, which have to be added coherently 
and correspond to the {\em two} diagrams in Fig.~\ref{fig1}.
In the case of the two-photon absorption described
by Eq.~(\ref{AsympApprox}), the two
diagrams analogous to Fig.~\ref{fig1} lead to equivalent
contributions, and only one term remains in the
matrix element.
In Eq.~(3), where necessary, the radiative corrections to the
energy levels of the intermediate states are included in the
denominators, according to the discussion of Ref.~\cite{Lo1952},
and are omitted where they are negligible.

We assume that the laser is tuned through the 
resonance $|i\rangle = |r\rangle$ near 
$\hbar \omega_{\mathrm L} \approx E_r - E_1$. 
The estimate of nonresonant contributions is determined as follows:
We define
\begin{equation}
\label{defx}
x = \hbar \omega_{\mathrm L} - (E_r - E_1)  
\end{equation}
as the (small) deviation from the resonance energy. The terms with 
$i = r$ in (\ref{TwoPhotonCross}) will give the dominant contribution.
All other intermediate states are
off resonance. Let $j \neq r$ denote the off-resonant states. Then we have
\begin{equation}
E_j - E_1 - \hbar \omega_{\mathrm L} = E_j - E_r - x\,,
\end{equation}
and the leading contributions to the cross section are given by
\begin{eqnarray}
\label{TwoPhotonNOR}
\frac{{\mathrm d}\sigma}{{\mathrm d}{\it \Omega}}
(\bbox{\varepsilon}_{\mathrm L}, 
\bbox{\varepsilon}_{\mathrm s}) 
 &\propto&
\frac{\left|
A_r(\bbox{\varepsilon}_{\mathrm L},\bbox{\varepsilon}_{\mathrm s})
  \right|^2}{x^2 + 
  \Gamma_r^2/4} 
+ \Bigg| \sum_{j\ne r}
  \frac{
A_j(\bbox{\varepsilon}_{\mathrm L},\bbox{\varepsilon}_{\mathrm s})}
  {E_j - E_r - x}\Bigg|^2
  - 2 \, {\rm Re}\Bigg\{ \Bigg[
\frac{ 
A_r(\bbox{\varepsilon}_{\mathrm L},\bbox{\varepsilon}_{\mathrm s})}
  {x + {\rm i}\,\Gamma_r/2} \Bigg]
\nonumber\\&&\times
  \Bigg[ \sum_{j\ne r}
  \frac{
  A_j^*(\bbox{\varepsilon}_{\mathrm L},\bbox{\varepsilon}_{\mathrm s})}
  { E_j - E_r - x}
+ \sum_{i}
  \frac{B_i^*(\bbox{\varepsilon}_{\mathrm L},\bbox{\varepsilon}_{\mathrm s})}
  {E_i + E_r - 2E_1 + x}\Bigg]
  \Bigg\} + \cdots
\end{eqnarray}
where
\begin{equation}
\label{defAi}
A_i(\bbox{\varepsilon}_{\mathrm L},\bbox{\varepsilon}_{\mathrm s})
= \sum_{\mu_i} \left(\bbox{\varepsilon}^*_{\mathrm s} 
  \cdot \bbox{D}_{1^\prime i}\right) \,
  \left(\bbox{\varepsilon}_{{\mathrm L}} \cdot
  \bbox{D}_{i1}\right)
\end{equation}
and
\begin{equation}
\label{defBi}
B_i(\bbox{\varepsilon}_{\mathrm L},\bbox{\varepsilon}_{\mathrm s})
= \sum_{\mu_i} \left(\bbox{\varepsilon}_{\mathrm L} 
  \cdot \bbox{D}_{1^\prime i}\right) \,
  \left(\bbox{\varepsilon}^*_{{\mathrm s}} \cdot
  \bbox{D}_{i1}\right)\,,
\end{equation}
where $\mu_i$ is the angular momentum projection of the state $| i \rangle$.
The right-hand side of Eq.~(\ref{TwoPhotonNOR}) contains the dominant resonance
in the first term.  The second term is only appreciable if the energy level 
of the state $j$ is separated from the resonance energy level by a 
fine-structure interval.  We neglect
the effects of hyperfine splitting in this discussion.  The third term
is the cross term between the resonant term and the other contributions, which
has been examined recently in Ref.~\cite{LaSoPlSo2001}.
The expression in (\ref{TwoPhotonNOR}) can be simplified by expanding in
powers of $x/(E_j-E_r)$ which is effectively 
an expansion in $\Gamma_r/(E_j-E_r)$
for $x$ of order $\Gamma_r$ or less.  Constant terms that contribute a flat
background and small corrections proportional to the resonance profile are 
omitted, with the result
\begin{eqnarray}
\label{TwoPhotonSimp}
\frac{{\mathrm d}\sigma}{{\mathrm d}{\it \Omega}}
(\bbox{\varepsilon}_{\mathrm L}, 
\bbox{\varepsilon}_{\mathrm s}) 
 &\propto&
\frac{\left|
A_r(\bbox{\varepsilon}_{\mathrm L},\bbox{\varepsilon}_{\mathrm s})
  \right|^2}{x^2 + 
  \Gamma_r^2/4} 
+ x\,\sum_{j\ne r} \sum_{k\ne r}
  \frac{A_j(\bbox{\varepsilon}_{\mathrm L},\bbox{\varepsilon}_{\mathrm s})\,
  A_k^*(\bbox{\varepsilon}_{\mathrm L},\bbox{\varepsilon}_{\mathrm s})}
  {(E_j - E_r)^2 \,(E_k - E_r)^2} \, \left(E_j + E_k -2\,E_r\right)
\nonumber\\[2ex]&&
  - 2 \, x \, {\rm Re}\Bigg\{
  \frac{
  A_r(\bbox{\varepsilon}_{\mathrm L},\bbox{\varepsilon}_{\mathrm s})}
  {x^2 + \Gamma_r^2/4} 
  \Bigg[ \sum_{j\ne r}
  \frac{
  A_j^*(\bbox{\varepsilon}_{\mathrm L},\bbox{\varepsilon}_{\mathrm s})}
  { E_j - E_r}
+ \sum_{i}
  \frac{B_i^*(\bbox{\varepsilon}_{\mathrm L},\bbox{\varepsilon}_{\mathrm s})}
  {E_i + E_r - 2E_1}\Bigg]
  \Bigg\} + \cdots\,.~~~
\end{eqnarray}
This expression gives a line shape which is a slightly shifted and
distorted Lorentz profile, that can be characterized as 
\begin{eqnarray}
\label{LineShape}
\frac{C}{x^2+\Gamma_r^2/4} + a\,x + \frac{b\,x}{x^2 + \Gamma_r^2/4} 
= \frac{C}{[x-\Delta(x)]^2+\Gamma_r^2/4} \,,
\end{eqnarray}
which, for $x$ of order $\Gamma_r$ or less, yields
\begin{eqnarray}
\Delta(x) = \frac{a}{2C}\left(x^2 + \Gamma_r^2/4\right)^2
                  + \frac{b}{2C}\left(x^2 + \Gamma_r^2/4\right) \,,
\end{eqnarray}
to lowest order in $\Delta(x)/\Gamma_r$.

We can now either
take the shift of the resonance curve at the half-maximum value
as the experimentally observable measure of the apparent shift
of the line center
\begin{eqnarray}
\label{approach1}
\Delta\left(\frac{\Gamma_r}{2}\right) = 
\frac{a\Gamma_r^4}{8C} + \frac{b\Gamma_r^2}{4C} \,,
\end{eqnarray}
or we can directly investigate the shift of the maximum of the resonance 
which is
\begin{eqnarray}
\label{approach2}
\Delta(0) = \frac{a\Gamma_r^4}{32C} + \frac{b\Gamma_r^2}{8C}
\end{eqnarray}
{\em Both} of these formulas 
are consistent with Low's measure of the asymmetry,
and indeed, the latter approach [the investigation of
$\Delta(0)$] was followed in~\cite{Lo1952}.
Formula (\ref{LineShape}) provides an estimate of the
line-shape distortion away from the maximum.

We assume that $\Gamma_r$ is of order $\alpha \, (Z\alpha)^4 \, mc^2$,
which is the case for hydrogenic P states, and take into account
the fact that the energy differences
are of order $(Z\alpha)^2 \, mc^2$
for states with different principle quantum numbers $n$, or are of
order $(Z\alpha)^4 \, mc^2$ for states with the same principle quantum
number. Furthermore, the following order-of-magnitude estimates
apply: $C \sim (Z\alpha)^{-4} (e \hbar/m c)^4$ (for hydrogenic
E1 transitions), $a \sim (Z\alpha)^{-16} (m c^2)^{-3} \,
(e \hbar/m c)^4$
[for energy numerators and denominators of order $(Z\alpha)^4 \, mc^2$],
and $b \sim (Z\alpha)^{-6} (m c^2)^{-1} \,
(e \hbar/m c)^4$ [for an energy denominator of order $(Z\alpha)^2 \, mc^2$].
We then we arrive immediately at the order-of-magnitude estimates for
$\Delta(0)/\Gamma_r$:
\begin{eqnarray}
\label{aorder}
\frac{a\Gamma_r^3}{32 C} \sim \left(\frac{\Gamma_r}{E_j-E_r}\right)^3
\sim \left\{\begin{array}{ll} \alpha^3 & \mathrm{for}~n_j = n_r
\\ & \\ \alpha^3\,(Z\alpha)^6  & \mathrm{for}~n_j \ne n_r
\end{array}   \right.
\end{eqnarray}
and
\begin{eqnarray}
\label{border}
\frac{b\Gamma_r}{8 C} \sim \frac{\Gamma_r}{E_j-E_r}
\sim \left\{\begin{array}{ll} \alpha & \mathrm{for}~n_j = n_r
\\ & \\ \alpha(Z\alpha)^2  & \mathrm{for}~n_j \ne n_r\,.
\end{array}   \right.
\end{eqnarray}
These order-of-magnitude estimates are not restricted to
the particular value $\Delta(0)$; they are valid for
the quantity $\Delta(x)/\Gamma_r$ for $x$ of the order of
$\Gamma_r$ or less. 
In particular, the estimates are relevant for the relative
shift of the peak of the resonance in a differential cross section
with respect to the natural line width. The contribution proportional to
$\alpha$ in the ``second term'' [see Eq.~(\ref{border})]
vanishes after summing over photon polarizations and
angular averaging over the emitted photons, i.e.~for the
total cross section; however it persists for the 
differential cross section, as it will be discussed
in Sec.~\ref{Differential}. In the second term,
contributions with $n_j = n_r$ can therefore be 
neglected for the total cross section. There are no further
cancellations in the total cross section (as compared to its differential form) 
for the other contributions given in Eqs.~(\ref{aorder})
and (\ref{border}).

These estimates can be illuminated by considering the particular example
of excitation from the 1S state to the 2P$_{1/2}$ state resonance
for hydrogen. In the case of a 1S--2P$_{1/2}$--1S transition, 
we have for the coefficient $C$ in the first term of
Eq.~(\ref{TwoPhotonSimp}) or (\ref{LineShape})
\begin{eqnarray}
\label{cexample}
C =
\frac{1}{2}\int {\rm d}{\it \Omega}
\sum_\mathrm{s}
\sum_{\mu,\,\mu^\prime}
\left| A_r(\bbox{\varepsilon}_{\mathrm L},
\bbox{\varepsilon}_{\mathrm s}) \right|^2
=\frac{2^{33}\mathrm{\pi}}{3^{22}(Z\alpha)^4}\left(\frac{e\hbar}{mc}\right)^4
+ \cdots \,,
\end{eqnarray}
where $\mathrm{d}{\it \Omega}$ is the solid angle element of the
emitted photon direction, $\mathrm{s}$ is the emitted photon polarization
index, $\mu$ and $\mu^\prime$ are the angular momentum projection of the
initial and final atomic
states, and a linearly polarized laser is assumed; 
the omitted terms indicated by dots are relativistic
corrections that are higher-order in $Z\alpha$.
For a general outline of the treatment of matrix elements of the type
(\ref{cexample}), we refer to~\cite{HiMo1980}\footnote{In this 
reference, the problem of electric field induced decay of the
2S state is examined within the QED scattering formulation.  The
formalism is very similar to the present considerations, because
the electric field acts as a zero frequency photon, which is very
far from resonance.}
. 
The prefactor $1/2$ in (\ref{cexample}) 
arises from the average over the initial state total angular
momentum projections.

The result is obtained by employing
relativistic wave functions.
Note that Eq.~(\ref{cexample}) is different if the electron
spin is neglected: in this case, the result should be multiplied 
by a factor 3. Specifically, the result obtained 
with ``spinless'' Schr\"{o}dinger wave functions is split in a
ratio $1/3$ to $2/3$ between the 1S--2P$_{1/2}$--1S and 
the 1S--2P$_{3/2}$--1S transitions, according to the different multiplicities
of the 2P states with different angular momentum. 
Laser spectroscopy can resolve the individual states, so we keep 
the spin in all intermediate stages of our calculations.

In order to investigate the off-resonant frequency shifts,
we turn our attention to the ``first term'' given in
Eq.~(\ref{aorder}). Its existence has been pointed out in~\cite{LaSoPlSo2002}.
Only terms for which both $j$ and $k$ correspond to
the 2P$_{3/2}$ state make an appreciable contribution. The coefficient
$a$ is given by
\begin{eqnarray}
\label{aexample}
a =
\int {\rm d}{\it \Omega}
\sum_\mathrm{s}
\sum_{\mu,\,\mu^\prime}
\frac{\left| A_j(\bbox{\varepsilon}_{\mathrm L},
\bbox{\varepsilon}_{\mathrm s}) \right|^2}{(E_j-E_r)^3}
=\frac{2^{50}\mathrm{\pi}}{3^{22}(Z\alpha)^{16}(mc^2)^3}
  \left(\frac{e\hbar}{mc}\right)^4
  + \cdots \,.
\end{eqnarray}
We use the well-known result
\begin{eqnarray}
\label{Gamma2P}
\Gamma_{2\mathrm{P}_{1/2}} = \frac{2^8}{3^8}\alpha(Z\alpha)^4\,mc^2
+\cdots\,,
\end{eqnarray}
which, together with Eqs.~(\ref{cexample}) and (\ref{aexample}),
yields
\begin{eqnarray}
\frac{a\Gamma_r^3}{32 C}
=\frac{2^{36}}{3^{24}}\,\alpha^3 + \cdots \approx 9.5\times10^{-8} \,,
\end{eqnarray}
corresponding to a frequency shift of 
about $9.5$~Hz.\footnote{In~\cite{LaSoPlSo2001}, a result of 4.89 Hz
is reported for this term.}

The ``second term'' given in Eq.~(\ref{border}) has been
investigated in 
Refs.~\cite{LaSoPlSo2001,LaSoPlSo2002}\footnote{In~\cite{LaSoPlSo2001}, 
a result of $+2.9$~Hz is reported for this term, whereas
in~\cite{LaSoPlSo2002}, the authors give an additional
factor of $4/9$ and a sign change to obtain a result of $-1.3$~Hz.
We agree with the magnitude of the former result and the sign of the
latter result.
Note that the term referred to as 
$\delta_{1{\mathrm S},1{\mathrm S}}^{2{\mathrm P}}$ in~\cite{LaSoPlSo2001}
corresponds to $\Delta(0)$ in our notation.}.
We obtain for this contribution ($b = b_1 + b_2$)
\begin{eqnarray}
\label{b1example}
b_1 &=& 
 - \int {\rm d}{\it \Omega}
\sum_\mathrm{s}
\sum_{\mu,\,\mu^\prime}
{\rm Re} \, A_r(\bbox{\varepsilon}_{\mathrm L},
\bbox{\varepsilon}_{\mathrm s})
\sum_{j\ne r}
\frac{A_j^*(\bbox{\varepsilon}_{\mathrm L},
\bbox{\varepsilon}_{\mathrm s})}{E_j-E_r}
= - \frac{\mathrm{9.3 \, \pi}}{(Z\alpha)^{6}\,mc^2}
  \left(\frac{e\hbar}{mc}\right)^4
  + \cdots \,,
\\[2ex]
\label{b2example}
b_2 &=& 
 - \int {\rm d}{\it \Omega}
\sum_\mathrm{s}
\sum_{\mu,\,\mu^\prime}
{\rm Re} \, A_r(\bbox{\varepsilon}_{\mathrm L},
\bbox{\varepsilon}_{\mathrm s})
\sum_i
\frac{B_i^*(\bbox{\varepsilon}_{\mathrm L},
\bbox{\varepsilon}_{\mathrm s})}{E_i+E_r-2E_1}
= - \frac{\mathrm{3.6 \, \pi}}{(Z\alpha)^{6}\,mc^2}
  \left(\frac{e\hbar}{mc}\right)^4
  + \cdots \,.
\end{eqnarray}
This yields for $Z=1$,
\begin{eqnarray}
\frac{b\Gamma_r}{8 C}
= -7.4 \times 10^{-2} \,
\alpha (Z\alpha)^2 + \cdots \approx -2.9 \times 10^{-8} \,,
\end{eqnarray}
corresponding to a frequency shift of $-2.9$~Hz.
Interference terms between the 2P$_{1/2}$ and 2P$_{3/2}$ states
vanish for $b_1$, but make an appreciable contribution to $b_2$.

In total, the frequency shift $\Delta(0)$ for the 
1S--2P$_{1/2}$--1S transition is
\begin{equation}
\label{resultP12}
\Delta(0) = (9.5 - 2.9)~\mbox{Hz} = +6.6~\mbox{Hz}
\qquad \mbox{(for 1S--2P$_{1/2}$--1S)}\,.
\end{equation}
This would be very difficult to detect
in comparison to the natural line width of about 100 MHz.

We now turn our attention to the four-photon process shown 
in Fig.~\ref{fig2}. From the usual $S$-matrix formalism we infer the 
cross-section to be proportional to
\begin{eqnarray}
\label{FourPhoton}
& & \frac{{\mathrm d}\sigma}{{\mathrm d}{\it \Omega}} \propto 
\nonumber\\
& & \!\!\!\!\!\!\!\! \left| \sum_{ijk}
\frac{\left(\bbox{\varepsilon}^*_{t} \cdot \bbox{D}_{1k}\right) \,
\left(\bbox{\varepsilon}^*_{\mathrm s} \cdot \bbox{D}_{kj}\right) \,
\left(\bbox{\varepsilon}_{{\mathrm L}} \cdot 
     \bbox{D}_{ji}\right) \,
\left(\bbox{\varepsilon}_{{\mathrm L}} \cdot 
     \bbox{D}_{i1}\right)}
  {\left[E_k - (E_1 + 2 \hbar \omega_{\mathrm L} - 
     \hbar \omega_{\mathrm s})\right] 
   \left[\left(E_j - \frac{\mathrm i}{2}\,\Gamma_j\right) - 
     \left(E_1 + 2 \hbar \omega_{\mathrm L}\right)\right]
   \left[E_i - (E_1 + \hbar \omega_{\mathrm L})\right]} + 
     \,\dots\, \right|^2 .
\end{eqnarray}
In this case, the energies of the two emitted photons 
$\omega_{\mathrm s}$ and $\omega_{\mathrm t}$ are subject to the condition
$\omega_{\mathrm s} + \omega_{\mathrm t} = 2 \omega_{\mathrm L}$. An integration
over one of the energies is required; this has no influence
on our considerations below. In this context,
it is probably worthwhile to note
that similar calculations involving four-photon processes
have to be performed in the context of 
third-harmonic generation (see \cite{Lo2000}).
Terms left out in the expression (\ref{FourPhoton}),
denoted by ``$\dots$'', correspond to the different 
time-orderings of photon emission and interactions with the 
laser [cf.~Eq.~(\ref{TwoPhoton})]. 
In the experiment, the laser is tuned through the 
two-photon resonance so that $E_1 + 2 \hbar \omega_{\mathrm L} \approx E_r$
where $E_r$ is the energy of the 2S state. In
the cross section (\ref{FourPhoton}), it is only the denominator
with summation variable $j$
which may become resonant, and which therefore is in need of a modification
$E_j \to E_j - \frac{\mathrm i}{2}\,\Gamma_j$. 
All other denominators in (\ref{FourPhoton}) remain 
off-resonant. 

For a Doppler-free 1S--2S two-photon process with the absorption
of two counter-propagating laser photons, 
a typical off-resonant contribution is given
by the case when $|j\rangle$ equals the 3S or 4S (in general, $n$S) state.
The matrix elements in the numerator of (\ref{FourPhoton}) have the same
order-of-magnitude for both the resonant and the off-resonant cases,
in analogy to the $C_{jr}$ being of order unity for the two-photon
process described by Eq.~(\ref{TwoPhotonNOR}).

We can now use exactly the same formalism as was used in the
analysis of the two-photon cross section (\ref{TwoPhotonCross}). 
A calculation shows that the nonresonant contributions to the four-photon
process described by Eq.~(\ref{FourPhoton}) result in a shift of the 
peak of the cross section by
\begin{equation}
\label{ApproxDeltaOmega2gamma}
\delta\omega_{2L} \sim \frac{\Gamma^2_r/\hbar}{E_j - E_r} \,,
\end{equation}
where $E_j$ denotes the energy of a typical nonresonant state.
The index $2L$ indicates that
there are {\em two} absorbed laser photons.

We will now consider the 1S--2S two-photon transition in 
atomic hydrogen.
The imaginary part of the energy (``the width'') of the 2S state is given
to within a good approximation by the 
two-photon radiative lifetime which is of the order of~\cite{MaMo1978}
\begin{equation}
\label{TwoPhotonWidth}
\Gamma_{\mathrm{2S}} \sim\alpha^2 \, (Z\alpha)^6\,m c^2 \,.
\end{equation}
This order-of-magnitude estimate can also be inferred 
by considering the poles and corresponding residues of
the low-energy part of the two-photon self-energy, as expressed
in nonrelativistic quantum electrodynamics [the relevant formula
is given in Eq.~(16) of~\cite{Pa2001}]. In view of 
Eqs.~(\ref{ApproxDeltaOmega2gamma}) and~(\ref{TwoPhotonWidth}),
\begin{equation}
\label{Result}
\delta\omega_{2L} \sim 
\alpha^4\,(Z\alpha)^{10} \,\frac{m c^2}{\hbar}\,.
\end{equation}
This is a factor of $\alpha^2\,(Z\alpha)^4$ smaller
than the width of the 2S state. For the 1S--2S transition in
atomic hydrogen, this estimate means that the off-resonant contributions
will enter at the level of $10^{-14}$ Hz. This is far smaller than the natural
width of the 2S state and irrelevant for current measurements
which have an experimental uncertainty of about~46~Hz~\cite{NiEtAl2000}.

In many cases, the shift of the peak of a resonance by off-resonant
contributions is much smaller than the natural width of the resonance
itself. This is not surprising and can be understood qualitatively.
A careful inspection of the physics associated with 
Eq.~(\ref{TwoPhotonSimp}) is sufficient. The derivative of the
narrow resonance contribution [first term of
Eq.~(\ref{TwoPhotonSimp})] near $x=0$ changes
rapidly in the vicinity of the resonance; by contrast, the 
off-resonant contributions [the remaining terms of Eq.~(\ref{TwoPhotonSimp})]
are rather flat in that frequency region. The maximum
of the shifted resonance profile must fulfill
the condition that the sum of the derivatives of resonant and off-resonant
contributions must add up to zero. The derivative of the 
resonant contribution changes rapidly, even {\em within} its 
natural width, and compensates that of the 
off-resonant contribution in the vicinity of the original resonant
contribution. This implies that the shifted peak of the resonance
must lie very close to its original value and can only be shifted
by a frequency difference which is much smaller than the natural width 
of the initial and final states in the resonance 
transition. 

Specifically, for the hydrogenic 1S--2S transitions,
there may be further nonresonant contributions, for example
those involving the absorption of {\em three} laser photons,
leading to an excitation of the 2S level into a P state
of the continuum with energy $E_{\mathrm{2S}} + \hbar \omega_{\mathrm L}$.
This further nonresonant effect will be proportional to the 
intensity of the laser beam. Because the transition 
matrix elements for bound state-continuum transition are enhanced
by a factor $(Z\alpha)^{-3/2}$ as compared to transitions
between two bound states, this effect will be of the order of 
$\alpha^4\,(Z\alpha)^7$. However, the order-of-magnitude
of this effect is still much smaller than the natural width
of the transition, in agreement with the qualitative discussion
in the preceding paragraph. 

Radiative vertex corrections to the photon scattering
also contribute to the line shape (see the discussion
in~\cite{Lo1952} and Fig.~\ref{fig3}).
So, our result in Eq.~(\ref{Result}) is not meant to indicate that the
peak of the resonance corresponds to the 1S--2S energy level difference
at the level of precision of $\delta \omega_{2L}$,
but it indicates that possibly interesting off-resonance effects
due to nearby atomic bound states are negligible for the 
two-photon high-precision experiments.

We conclude this section by noting that there is a further
nonresonant contribution originating from the possibility of
the nonrelativistically forbidden, nonresonant {\em one-photon}
magnetic 1S--2S transition. This contribution is, however, Doppler-broadened 
[cf.~Eq.~(\ref{AsympApprox})] and is 
therefore negligible to a very good approximation. 
Its magnitude will depend on the experimental conditions.
Also, it is known that the {\em second-order Doppler effect} and 
the {\em AC Stark shifts} in the
intense laser field are the most important
systematic effects in current experiments~\cite{NiEtAl2000} (these
necessitate an extrapolation to zero laser field).

%
%
\section{Differential measurement}
\label{Differential}

We consider the process 1S--$2{\mathrm P}_{3/2}$--1S
described by the cross section (\ref{TwoPhotonCross}).
There is a nearby $2{\mathrm P}_{1/2}$ level 
which is separated from the $2{\mathrm P}_{3/2}$-level
by the fine-structure interval
\begin{equation}
E(2{\mathrm P}_{3/2}) - E(2{\mathrm P}_{1/2}) = 
  \frac{(Z\alpha)^4}{32}\,mc^2.
\end{equation}
This is two orders of $(Z\alpha)$ smaller than the hydrogenic 
energy difference between states with different quantum numbers,
which is of order $(Z\alpha)^2$. According to 
Eq.~(\ref{border}),
the nearby $2{\mathrm P}_{3/2}$ level should lead to a large
nonresonant contribution of order
\begin{equation}
\label{ApproxFS}
\delta \omega(1{\mathrm S} \to 2{\mathrm P}_{3/2} \to 1 {\mathrm S}) \sim 
\frac{(\Gamma_{2{\mathrm P}_{1/2}})^2/\hbar}
  {E(2{\mathrm P}_{3/2}) - E(2{\mathrm P}_{1/2})} \sim
     \alpha^2 \, \frac{(Z \alpha)^8}{(Z\alpha)^4}\frac{mc^2}{\hbar} =
      \alpha^2 \, (Z\alpha)^4 \frac{mc^2}{\hbar}.
\end{equation}
which is two orders of $(Z\alpha)^2$ larger\footnote{Some arguments 
presented in the seminal paper~\cite{Lo1952}
with regard to nonresonant levels which are ``removed from 
the (intermediate state) $m$ (of the two-photon process)
by a fine- or hyperfine-structure
splitting'' do not appear to be universally applicable 
[it was argued that the smaller energy denominator 
in this case is compensated by smaller transition matrix
elements $\bbox{D}_{m1}$ in the numerator of~(\ref{TwoPhotonCross})].
This argument is certainly applicable
to off-resonant states which lie very close
to the {\em initial} state of the two-photon process.
However, in the presence of an {\em intermediate} state
very close to the resonance state, whose transition matrix
elements are of the same order as for the resonant state,
this argument does not appear to hold universally. 
In this case, a treatment
including fine structure effects cannot be avoided,
especially if differential cross sections are considered.}
than the result in Eq.~(\ref{resultP12}).
Based on Eqs.~(\ref{border}), (\ref{Gamma2P}) and
(\ref{b1example}), we obtain for the 
nonresonant frequency shift of the process 1S--$2{\mathrm P}_{3/2}$--1S
the result
\begin{equation}
\label{ResultFS}
\delta \omega(1{\mathrm S} \to 2{\mathrm P}_{3/2} \to 1 {\mathrm S}) =  
K \, \left[ 2^3 \left(\frac{2}{3}\right)^{16} \, \alpha^2 \,
  (Z\alpha)^4 \, \frac{m c^2}{\hbar} \right]\,,
\end{equation}
where $K$ is a numerical factor which depends on spin and photon polarizations.
When summing over the final-state
spin polarizations of the electron, summing over the photon polarizations
{\em and} integrating over the angle of the emitted photon,
$K$ vanishes. In this case, the dominant 
off-resonant frequency 
is of order $\alpha^2\,(Z\alpha)^6mc^2/\hbar$~[see Eq.~(\ref{LowEstim})
and Refs.~\cite{Lo1952,LaSoPlSo2001}]. However, the coefficient
$K$ does {\em not} vanish when spin-polarized hydrogen is
used and measured in the experiment, or differential cross sections
are considered. 

For the
transition 1S($m=\lfrac{1}{2}$)--$2{\mathrm P}_{3/2}$--1S($m=\lfrac{1}{2}$),
there is a nonvanishing interference term which depends on
the polarization of the laser beam (here, $m$ refers
to the projection of the electron spin onto the $z$ axis). Specifically,
$K = 1/2$ for linearly polarized light along the $z$ axis,
and $K = 2$ for $\sigma^{-}$-polarized light.
Here, $\sigma^{-}$-polarized light is to be understood as
circularly polarized light with a polarization vector in the
$x$--$y$ plane; this polarization being chosen such that the light
can drive both transitions 1S($m=\lfrac{1}{2}$)--$2{\mathrm P}_{3/2}$ 
and 1S($m=\lfrac{1}{2}$)--$2{\mathrm P}_{1/2}$.
For $\sigma^{+}$-polarized light, which can 
drive a transitions 1S($m=\lfrac{1}{2}$)--$2{\mathrm P}_{3/2}$ 
but cannot drive 1S($m=\lfrac{1}{2}$)--$2{\mathrm P}_{1/2}$,
$K$ again vanishes.
The average ${\overline K} = 5/4$ for the two experimental set-ups for
which $K$ is nonvanishing corresponds to a shift of order ${\delta \nu} = 
\delta \omega/(2 \pi) = 0.28~{\mathrm{MHz}}$.
This could be within
the range of current measurements, even though it has to be
compared to the natural width of the hydrogenic 2P states which 
is of the order of $\Gamma(2{\mathrm P}) = 100~{\mathrm{MHz}}$.

We also note that $K$ is nonvanishing if the experiment is restricted
to the measurement of specific photon polarizations. For example,
for the case of an initial 1S($m=\lfrac{1}{2}$) state, with a laser
beam polarized along the $z$ axis, and emitted
photon polarizations along either
the $x$ or the $y$ axis, we obtain $K = -1$, corresponding to
$\delta \omega/(2 \pi) = -0.23~{\mathrm{MHz}}$.
Also, with a both the laser beam and the emitted
photons polarized along the $z$ axis, we obtain $K = 1/2$,
corresponding to $\delta \omega/(2 \pi) = +0.11~{\mathrm{MHz}}$.

The obvious idea would be to switch ``on or off''
the nonresonant contribution from the nearby $2{\mathrm P}_{1/2}$
state. This can be reached by switching from a measurement
involving unpolarized light and/or spin-unpolarized hydrogen to a
measurement with polarized light and/or spin-polarized atomic
hydrogen, as described above. 

%
%
\section{Conclusions}
\label{Conclusions}

We have analyzed nonresonant contributions to two-photon and 
four-photon processes shown in Figs.~\ref{fig1} and~\ref{fig2}.
We provide a formula for the distorted Lorentz profile
in Eq.~(\ref{LineShape}).
For the hydrogenic 1S--2P$_{1/2}$ transitions, the result for
the off-resonant frequency shifts is given in Eq.~(\ref{resultP12}).

For the hydrogenic 1S--2S transition, we obtain the order-of-magnitude
estimate (\ref{ApproxDeltaOmega2gamma}) for 
the nonresonant corrections to the observed transition 
frequency as one of the main results of this paper. 
The nonresonant effect due to off-resonant atomic bound states
is given in Eq.~(\ref{Result}) and is negligible on
the level of any conceivable measurement (see Sec.~\ref{TwoPhoton}). 
Notably, this nonresonant correction is orders of magnitude
smaller than other systematic effects like the second-order Doppler effect and the 
AC Stark shift. However, three-photon
transitions to the continuum and relativistic and 
radiative corrections to the two-photon process
(a typical diagram is given in Fig.~\ref{fig3}) 
will dominate the off-resonant frequency shifts for two-photon transitions
(see the discussion in Sec.~\ref{TwoPhoton}),
and these and other effects can be orders of magnitude larger than 
the effect due to off-resonant S levels given in Eq.~(\ref{Result})
(see also the related discussion in~\cite{LaSoPlSo2002}).
While the magnitude of these effects depends on the 
particular experimental conditions (e.g., the laser intensity
or the measurement of differential vs.~total cross sections),
we conclude that the natural line shape of the two-photon
process will remain Lorentzian to a very good approximation,
enabling the comparison of theory and experiment for the 
1S--2S transition to well below the 1.3~Hz natural line width.

We discuss the possibility of a differential measurement
by which a nonresonant contribution from a nearby state
could be detected in current Lamb shift measurements 
(see Sec.~\ref{Differential}). For the specific experimental setup 
outlined in Sec.~\ref{Differential}, 
the requirement would be to determine the line center -- defined
as the frequency corresponding to the maximum cross section -- to within
a relative uncertainty of the order of one part in 400; 
this measurement could be feasible with the current technology.
The nonresonant shift in this case is indicated in Eq.~(\ref{ResultFS}).
Clearly, an enhancement of nonresonant contributions can be expected in
an atomic system where two levels with equal quantum numbers are
lying very close to each other.

The discussion in Sec.~\ref{Differential} shows that 
nonresonant levels which lie close to a resonance
may result in surprisingly large shifts of the peaks of the 
cross sections. However, the qualitative considerations presented
near the end of Sec.~\ref{TwoPhoton} always remain valid:
the ``nonresonant shift'' $\delta \omega$
constitutes only a fraction of the natural radiative 
width of the resonance if the natural width of the transition
is small compared to the energy interval 
between the resonant and the nearest off-resonant state.
This consideration also applies if levels lie very close to
each other, and the denominator in Eq.~(\ref{ApproxDeltaOmega2gamma}) or
(\ref{ApproxFS}) becomes very small:
one example would be hyperfine-singlet and
hyperfine-triplet hydrogenic S levels ($F=0,1$) which are
separated from each other only by the
hyperfine structure splitting.

%
%
\section*{Acknowledgements}

UDJ is grateful for continued support extended by NIST during
a series of research appointments, and to D. Kelleher,
K. Pachucki and V. M. Shabaev for helpful discussions.


\begin{thebibliography}{10}

\bibitem{NiEtAl2000}
M. Niering, R. Holzwarth, J. Reichert, P. Pokasov, T. Udem, M. Weitz, T.~W.
  H\"{a}nsch, P. Lemonde, G. Santarelli, M. Abgrall, P. Laurent, C. Salomon,
  and A. Clairon, Phys. Rev. Lett. {\bf 84},  5496  (2000).

\bibitem{JeMoSo1999}
U.~D. Jentschura, P.~J. Mohr, and G. Soff, Phys. Rev. Lett. {\bf 82},  53
  (1999).

\bibitem{MeRi2000}
K. Melnikov and T. v.~Ritbergen, Phys. Rev. Lett. {\bf 84},  1673  (2000).

\bibitem{Ye2000}
V.~A. Yerokhin, Phys. Rev. A {\bf 62},  012508  (2000).

\bibitem{JeMoSo2001pra}
U.~D. Jentschura, P.~J. Mohr, and G. Soff, Phys. Rev. A {\bf 64},  042512
  (2001).

\bibitem{MoTa2000}
P.~J. Mohr and B.~N. Taylor, Rev. Mod. Phys. {\bf 72},  351  (2000).

\bibitem{KrHe1925}
W. Kramers and W.~H. Heisenberg, Z. Phys. {\bf 31},  681  (1925).

\bibitem{Lo2000}
R. Loudon, {\em The Quantum Theory of Light}, third ed. (Oxford University
  Press, 2000).

\bibitem{Lo1952}
F. Low, Phys. Rev. {\bf 88},  53  (1952).

\bibitem{Bi1995}
C. Billionet, J. Phys. I (France) {\bf 5},  949  (1995).

\bibitem{LundeenThesis}
S.~R. Lundeen, PhD thesis, Harvard University, 1981 (unpublished).

\bibitem{La1983}
L.~N. Labzowsky, Zh. \'{E}ksp. Teor. Fiz. {\bf 85},  869  (1983).

\bibitem{LaKlDm1993}
L.~N. Labzowsky, G. Klimchitskaya, and Y. Dmitiriev, {\em Relativistic Effects
  in the Spectra of Atomic Systems} (IoP, Bristol, 1993).

\bibitem{KaEtAl1992}
V.~V. Karasiev, L.~N. Labzowsky, A.~V. Nefiodov, V.~G. Gorshkov, and A.~A.
  Sultanaev, Phys. Scr. {\bf 46},  225  (1992).

\bibitem{LaEtAl1993}
L.~N. Labzowsky, V.~V. Karasiev, I. Lindgren, H. Persson, and S. Salomonson,
  Phys. Scr. T {\bf 46},  150  (1993).

\bibitem{LaKaGo1994}
L.~N. Labzowsky, V.~V. Karasiev, and I.~A. Goidenko, J. Phys. B {\bf 27},  L439
   (1994).

\bibitem{LaSoPlSo2001}
L.~N. Labzowsky, D.~A. Solovyev, G. Plunien, and G. Soff, Phys. Rev. Lett. {\bf
  87},  143003  (2001).

\bibitem{HiMo1980}
M. Hillery and P. J. Mohr, Phys. Rev. A {\bf 21}, 24 (1980).

\bibitem{LaSoPlSo2002}
L.~N. Labzowsky, D.~A. Solovyev, G. Plunien, and G. Soff, e-print
  physics/0201067, version 1 of 30 Jan, 2002.

\bibitem{JeSoMo1997}
U.~D. Jentschura, G. Soff, and P.~J. Mohr, Phys. Rev. A {\bf 56},  1739
  (1997).

\bibitem{GM1931}
M. G\"{o}ppert-Mayer, Ann. Phys. (N. Y.) {\bf 9},  273  (1931).

\bibitem{KaGa1961}
W. Kaiser and C.~G. Garret, Phys. Rev. Lett. {\bf 7},  229  (1961).

\bibitem{De1993}
W. Demtr\"{o}der, {\em Laserspektroskopie}, 3 ed. (Springer, Berlin, 1993).

\bibitem{MaMo1978}
R. Marrus and P. J. Mohr, Adv. At. Mol. Phys. {\bf 14},
  181 (1978).

\bibitem{Pa2001}
K. Pachucki, Phys. Rev. A {\bf 63},  042503  (2001).

\end{thebibliography}
\end{document}